\documentclass[useAMS,usenatbib]{mn2e}

\usepackage{amsmath, amssymb}
\usepackage[dvips]{graphicx}
\usepackage{epsfig}

\newcommand{\rlight}{r_{\rm L}}

\newcommand{\ex}{\vec{e}_{\rm x}}
\newcommand{\ey}{\vec{e}_{\rm y}}
\newcommand{\ez}{\vec{e}_{\rm z}}

\newcommand{\aap}{A\&A}

\newcommand{\apj}{ApJ}
\newcommand{\apjl}{ApJL}
\newcommand{\apjs}{ApJS}
\newcommand{\aplett}{Astrophysical Letters}

\title[An unified polar cap/striped wind model for pulsars]{An unified polar
  cap/striped wind model for pulsed radio and gamma-ray emission in
  pulsars} \author[J. P\'etri]{J. P\'etri$^{1}$\thanks{E-mail:
    jerome.petri@astro.unistra.fr}\\
  $^{1}$Observatoire Astronomique de Strasbourg, Universit\'e de
  Strasbourg, CNRS, UMR 7550, 11 rue de l'Universit\'e, 67000
  Strasbourg, France.}

\begin{document}

\date{Accepted . Received ; in original form }

\pagerange{\pageref{firstpage}--\pageref{lastpage}} \pubyear{2010}

\maketitle

\label{firstpage}

\begin{abstract} 
  Thanks to the recent discovery by Fermi of about fifty new gamma-ray
  pulsars, it becomes possible to look for statistical properties of
  their pulsed high-energy emission, especially their light-curves and
  phase-resolved spectra.  These pulsars emit by definition mostly
  gamma-ray photons but some of them are also detected in the radio
  band. For those seen in these two extreme energies, the relation
  between time lag of radio/gamma-ray pulses and gamma-ray peak
  separation, in case both high-energy pulses are seen, helps to put
  some constrain on the magnetospheric emission mechanisms and
  location.  This idea is analyzed in detail in this paper, assuming a
  polar cap model for the radio pulses and the striped wind geometry
  for the pulsed high-energy counterpart.
  
  Combining the time-dependent emissivity in the wind, supposed to be
  inverse Compton radiation, with a simple polar cap emission model
  along and around the magnetic axis, we compute the radio and
  gamma-ray light-curves, summarizing the results in several phase
  plots.  The phase lag as well as the gamma-ray peak separation
  dependence on the pulsar inclination angle and on the viewing angle
  are studied.  Using the gamma-ray pulsar catalog compiled from the
  Fermi data, we are able to predict the radio lag/peak separation
  relation and compare it with available observations taken from this
  catalog.

  This simple geometric model, combining polar cap and striped wind
  radiation performs satisfactorily at explaining the observed
  radio/gamma-ray correlation. This supports the idea of distinct
  emission locations for the radio and gamma-ray radiation.
  Nevertheless, time retardation effects like curved-space time and
  magnetic field lines winding up close to the neutron star can lead
  to discrepancy between our predicted time lag and a more realistic
  relation as deduced from the gamma-ray catalog. Moreover, no
  accurate polar cap description being at hand so far, large
  uncertainties remains on the altitude and geometry of the radio
  emission.
\end{abstract}

\begin{keywords}
  Pulsars: general - Radiation mechanisms: non-thermal - Gamma rays:
  observations - Gamma rays: theory - Stars: winds, outflows
\end{keywords}

\section{INTRODUCTION}

The new catalog on gamma-ray pulsars obtained by the Fermi-LAT
instrument \citep{2010ApJS..187..460A} increased the number of
gamma-ray pulsars from seven to about fifty. Since then, new pulsars
are discovered regularly. 
This allows for the first time a reasonable statistical analysis on
the high-energy emission properties of these objects like spectral
shape, cut-off energy, and comparison between radio and gamma-ray
radiation if available.  For some high luminosity pulsars, this
analysis has been completed by a phase-resolved study like for the
Crab \citep{2010ApJ...708.1254A}, Vela \citep{2009ApJ...696.1084A} and
Geminga \citep{2010ApJ...720..272A}.  Radio pulses and gamma-ray
photons are expected to be produced in different emission sites,
probably close to the neutron star surface for the former, described
by a polar cap model \citep{1969ApL.....3..225R}, and in the vicinity
of the light-cylinder for the latter, explained by outer gaps
\citep{1986ApJ...300..500C} or alternatively by the striped wind
\citep{2009A&A...503...13P}. Polarization properties at
multi-wavelength would certainly help to constrain the geometry of the
sites of emission \citep{2004ApJ...606.1125D, 2005ApJ...627L..37P,
  2009A&A...503...13P}.  Gamma-ray light-curves alone can already give
good insight into the magnetosphere \citep{2010ApJ...714..810R}.

The new sample of Fermi-LAT gamma-ray pulsars increased interest into
modeling of gamma-ray emission. Detection of several millisecond
gamma-ray pulsars was not expected and came as a real surprise.  Thus,
\cite{2009ApJ...707..800V} focused special attention to this class of
millisecond pulsars to probe the geometry of the emission regions,
taking into account relativistic effects.

Very recently, gamma-ray light-curves have been computed for the
simple vacuum dipole model \citep{2010ApJ...715.1270B} or better for a
realistic magnetospheric model based on 3D MHD simulations of the near
pulsar magnetosphere. This requires some post-processing prescription
about the emission location and mechanism within the magnetosphere
\citep{2010ApJ...715.1282B}. In this model, gamma rays are expected
close to the light-cylinder.

An alternative site for the production of pulsed radiation has been
investigated a few years ago by~\cite{2002A&A...388L..29K}. This model
is based on the striped pulsar wind, originally introduced by
\cite{1990ApJ...349..538C} and \cite{1994ApJ...431..397M}.  Emission
from the striped wind originates outside the light cylinder and
relativistic beaming effects are responsible for the phase coherence
of this radiation. It has already been shown that this model can
satisfactorily fit the optical polarization data from the Crab
pulsar~\citep{2005ApJ...627L..37P} as well as the phase-resolved
high-energy spectral variability of Geminga
\citep{2009A&A...503...13P}.

The aim of this work is to complete the atlas of light-curves and
phase plots performed by \cite{2009ApJ...695.1289W} on the ground of
the polar cap, the slot gap and the outer gap models. It furnishes a
extended benchmark to test different scenarios for the physical
processes occurring in the pulsar magnetosphere.  We use an explicit
asymptotic solution for the large-scale magnetic field structure
related to the oblique split monopole~\citep{1999A&A...349.1017B} and
responsible for the gamma-ray light-curves combined with a simple
polar cap geometry for the radio counterpart. Details are given in
Sec.~\ref{sec:Modele}. We then compute the properties of the radio and
high-energy light curves for the pulsed emission and compare our
results with several gamma-ray pulsars extracted from the Fermi
catalog.  Relevant results are discussed in Sec.~\ref{sec:Results}
before concluding this study.

\section{THE MODEL}
\label{sec:Modele}

The model employed to compute the gamma-ray pulse shape emanating from
the striped wind is briefly reexamined in this section. The
geometrical configuration and emitting particle distribution functions
follows the same lines as those described in
\cite{2009A&A...503...13P}.  The magnetized neutron star rotates at an
angular speed of~$\Omega_*$ (period~$P=2\,\pi/\Omega_*$) directed
along the $(Oz)$-axis, i.e. the rotation axis is given by
$\vec{\Omega}_*= \Omega_* \, \ez$. We use a Cartesian coordinate
system with coordinates~$(x,y,z)$ and orthonormal basis~$(\ex, \ey,
\ez)$. The stellar magnetic moment is denoted by~$\vec{\mu}_*$, it is
assumed to be dipolar and makes an angle~$\chi$ with respect to the
rotation axis such that
\begin{equation}
  \label{eq:momamg}
  \vec{\mu}_* = \mu_* \, [ \sin\chi \, ( \cos (\Omega_* \, t) \, \ex + 
  \sin (\Omega_* \, t) \, \ey ) + \cos\chi \, \ez ] .
\end{equation}
This angle is therefore defined by $\cos\chi = \vec{\mu}_* \cdot \ez /
\mu_*$. The inclination of the line of sight with respect to the
rotational axis, and defined by the unit vector $\vec{n}$, is denoted
by~$\zeta$.  It lies on the $(Oyz)$~plane, thus
\begin{equation}
  \vec{n} = \sin\zeta \, \ey + \cos\zeta \, \ez .
\end{equation}
Thus we have $\cos\zeta = \vec{n} \cdot \ez$. Moreover, the wind
expands radially outwards at a constant velocity~$V$ close to the
speed of light denoted by~$c$.

Because we are not interested in the phase-resolved spectra, but only
in the light-curves above 100~MeV, our model only involves geometrical
properties related to the magnetic field structure and viewing angle.
The particle distribution function does not play any role except for
its density number. Dynamical properties (such that energy
distribution) related to the emitting particles are unimportant in
this case. In order to compute the light curves, we use a simple
expression for the emissivity of the wind related to inverse Compton
scattering.  This is explained in the next paragraphs.

\subsection{Magnetic field structure}

The geometrical structure of the wind is based on the asymptotic magnetic
field solution given by \cite{1999A&A...349.1017B}.  Outside the light
cylinder, the magnetic structure is replaced by two magnetic monopoles
with equal and opposite intensity. The current sheet sustaining the
magnetic polarity reversal arising in this solution, expressed in
spherical coordinates~$(r, \vartheta, \varphi)$ is defined by
\begin{equation}
  \label{eq:Rs}
  r_{\rm s}(\vartheta,\varphi,t) = \beta_{\rm v} \, \rlight \, \left[ 
    \pm \arccos ( - \cot\vartheta \, \cot\chi) + \frac{c\,t}{\rlight} - 
    \varphi + 2\,l\,\pi \right]
\end{equation}
where $\beta_{\rm v} = V/c$, $\rlight=c/\Omega_*$ is the radius of the
light cylinder, $t$ is the time as measured by a distant observer at
rest, and $l$ an integer. Because of the ideal MHD assumption, this
surface is frozen into the plasma and therefore also moves radially
outwards at a speed~$V$.  Strictly speaking, the current sheets are
infinitely thin and the pulse width would be inversely proportional to
the wind Lorentz factor~$\Gamma_{\rm v} = (1-\beta_{\rm v}^2)^{-1/2}$,
\citep{2009ASSL..357..421K}. Here, as already done for the study of
the synchrotron polarization of the pulsed emission
\citep{2005ApJ...627L..37P} and for Geminga
\citep{2009A&A...503...13P} we release this restrictive and unphysical
prescription. Indeed, the current sheets are assumed to have a given
thickness, parameterized by the quantity~$\Delta_\varphi$ (see
Eq.~(\ref{eq:Epaisseur}) below for an explicit expression of the
thickness).  Moreover, inside the sheets, the particle number density
is very high while the magnetic field is weak.  In whole space, the
magnetic field is purely toroidal and given by
\begin{equation}
  \label{eq:Bphi}
  B_\varphi = B_{\rm L} \, \frac{R_{\rm L}}{r} \, \eta_\varphi
\end{equation}
The strength of the magnetic field at the light-cylinder is denoted
by~$B_{\rm L}$. In the original work of \cite{1999A&A...349.1017B} ,
the function $\eta_\varphi$ is related to the Heaviside unit step
function and can only have two values $\pm 1$, leading to the
discontinuity in magnetic field. In order to make the transition more
smooth, we redefine the function~$\eta_\varphi$ by
\begin{eqnarray}
  \label{eq:etaphi}
  \eta_\varphi & = & \tanh( \Delta_\varphi \, \psi ) \\
  \psi & = & \cos \vartheta \, \cos \chi + \sin \vartheta \, \sin \chi \, 
  \cos\left[ \varphi - \Omega_* \, ( t - \frac{r}{V} ) \right] \nonumber
\end{eqnarray}
With these formulae, the physical length of the transition layer has a
thickness of the order of
\begin{equation}
  \label{eq:Epaisseur}
  2\,\pi\,\beta_{\rm v}\,\rlight/\Delta_\varphi .
\end{equation}

\subsection{Particle density number}

The aim of this paper is to show the behavior of the pulsed
high-energy light-curves emanating from the striped wind flow for
different magnetic plus emitting particle configurations. We will not
perform a detailed study of the phase-resolved spectral variability.
Thus for this purpose, it is sufficient to fix the particle density
number without specifying the distribution in momentum space. We will
assume a $e^\pm$ pair mixture flowing dominantly within the current
sheets defined by Eq.~(\ref{eq:Rs}).  We thus adopt the following
expression for the total particle density number
\begin{equation}
  \label{eq:Densite}
  K_e(\vec{r},t) = \frac{( N - N_0 ) \, 
    {\rm sech}^2 (\Delta_\varphi \, \psi) + N_0}{r^2}
\end{equation}
$N_0$ sets the minimum density in the stripes, between the current
sheets, whereas $N$ defines the highest density inside the sheets.  We
refer to \cite{2009A&A...503...13P} for more details about the
justification of this choice.

\subsection{Inverse Compton emissivity in the wind}

These particles will be visible through their inverse Compton
radiation.  The corresponding total emissivity is denoted by
$j_\mathrm{ic}^\mathrm{obs}$. Strictly speaking, this radiated power
should depend on space location $\vec r$, time of observation~$t$ as
well as on the energy of the detected photons. However, recall that a
spectral study is out of the scope of this work, so the energy
dependence can be dropped. We only retain variation with $\vec r$ and
$t$.

As done in \cite{2009A&A...503...13P}, the inverse Compton light
curves are obtained by integrating this emissivity over the whole wind
region. This wind is assumed to extend from a radius~$r_0$ to an outer
radius~$r_{\rm s}$ which can be interpreted as the location of the
termination shock. Therefore, the inverse Compton radiation at a fixed
observer time~$t$ is given by
\begin{eqnarray}
  \label{eq:CourbeLum}
  I_\mathrm{ic}^\mathrm{obs}(t) & = & \!\!\!
  \int_{r_0}^{r_{\rm s}} \!\!\! \int_{0}^{\pi} \!\!\! \int_0^{2\pi}
  \!\!\! j_\mathrm{ic}^\mathrm{obs}(\vec{r}, t_{\rm ret}) \, r^2  
  \sin\vartheta dr d\vartheta d\varphi
\end{eqnarray}
The retarded time is expressed as $t_{\rm ret} = t - ||\vec{R}_0 -
\vec{r}||/c \approx t - R_0/c + \vec{n} \cdot \vec{r} / c$.
Eq.(\ref{eq:CourbeLum}) is integrated numerically. We compute the
inverse Compton intensity for several geometries. We are therefore
able to predict the phase resolved pulse shape, for any inclination of
the line of sight and any obliquity of the pulsar.  Results and
applications to some $\gamma$-ray pulsars are discussed in the next
section.

\subsection{Polar cap emissivity}

The polar caps or their close vicinity to an altitude up to a few
stellar radii from the stellar crust are the favorite sites to explain
the coherent pulsed radio emission. For our geometric model, both the
polar caps radiate most efficiently at their center which means along
the magnetic axis. When moving away from the poles staying on the
stellar surface, the radio intensity should decrease. We therefore
choose a smooth gaussian decreasing with distance from the magnetic
north or south pole, almost complete extinction occurring outside the
polar caps. Their size is determined by the location of the foot of
the last open magnetic field lines, on the stellar surface. For
simplicity, neither bending of magnetic field lines nor
(general-)relativistic effects nor plasma effects are included. The
distortions implied by these effects are not included in our
description. Moreover, using the expression for the aligned rotator,
the size of one polar cap, assumed to have a circular shape, is
obtained from its angular opening angle $\vartheta_{\rm pc}$ leading
to a radius $R_{\rm pc}$ such that
\begin{eqnarray}
  \label{eq:DimensionCalotte}
  \sin\vartheta_{\rm pc} & = & \sqrt{\frac{R_*}{\rlight}} \\
  R_{\rm pc} & = & R_* \, \vartheta_{\rm pc}
\end{eqnarray}
The maximal intensity arises when line of sight and magnetic moment
are aligned or counter-aligned. For a sharp transition between on and
off states, emission is expected only when the angle formed by
$\vec\mu_*$ and $\vec n$ is less than the angular opening angle, thus
$\widehat{(\vec\mu_*,\vec n)} \le \vartheta_{\rm pc}$. In other words, a polar
cap (either north or south) is only seen if
\begin{equation}
  \label{eq:CalotteVisible}
  \cos \, (\vec\mu_*,\vec n) = \frac{\vec\mu_* \cdot \vec n}{\mu_*} \ge \cos \vartheta_{\rm pc} = \sqrt{1-\frac{R_*}{\rlight}}
\end{equation}
This is translated mathematically for our gaussian by introducing the
intermediate function $\Phi_{\rm pc}$ defined as
\begin{equation}
  \label{eq:EmissionCalotte1}
  \Phi_{\rm pc}(t) = \frac{\vec\mu_* \cdot \vec n}{\mu_*} = \cos \zeta \, \cos \chi + \sin \zeta \, \sin \chi \, \sin (\Omega_*\,t)
\end{equation}
Therefore one polar cap is visible if
\begin{equation}
  \label{eq:EmissionCalotte2}
  \Phi_{\rm pc}^2(t) \ge 1 - \frac{R_*}{\rlight}
\end{equation}
Expressed as a polar cap emissivity $\epsilon_{\rm pc}$ with gaussian
similar shape, we write it
\begin{equation}
  \epsilon_{\rm pc} \propto e^{\alpha \, (\Phi_{\rm pc}^2 - 1 + R_*/\rlight)}
\end{equation}
with $\alpha$ a positive constant controlling the extension of
significant polar cap emission. $\epsilon_{\rm pc}$ is maximum for
$\vec\mu_*$ and $\vec n$ colinear with value close to unity and tends
to zero for large distance from the magnetic poles.  We stress that
this description is purely phenomenological, to be included in the
phase plot for gamma-ray light-curves.

\section{RESULTS}
\label{sec:Results}

Our simple geometrical model allows to perform some analytical
calculations to deduce the relative lag in arrival time between radio
and gamma-ray photons and the gamma-ray peak separation, if both,
radio emission and double peak structure are detectable. In the
following paragraphs, we perform a detailed analysis of the
light-curve properties and correlation with radio wave-bands depending
on $\chi$ and $\zeta$. We then close the discussion with application
to some gamma-ray pulsars from the Fermi first year catalog.

\subsection{The high-energy peak separation}

\begin{figure}
  \centering
  \includegraphics[width=0.45\textwidth]{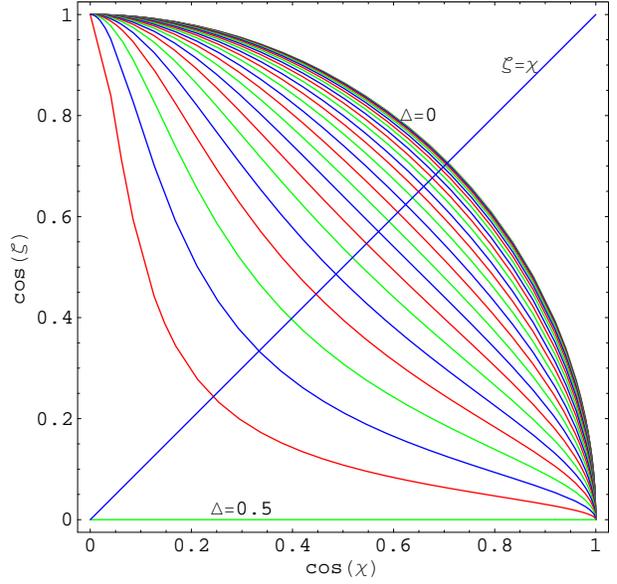} 
  \caption{Peak separation $\Delta$ between both gamma-ray pulses
    purely determined by the obliquity of the pulsar~$\chi$ and the
    inclination angle of the observer line of sight~$\zeta$. Each
    curve represents a constant value of $\Delta$ ranging from 0 to
    0.5 by 0.02~steps. No separation occurs along the circle of radius
    unity (pulse overlapping) whereas maximum separation of 0.5 is
    reached for $\cos\zeta=0$.  The diagonal $\zeta=\chi$ describes
    the visibility of one magnetic pole.}
  \label{fig:SeparationPic}
\end{figure}

In the striped wind model, the light-curve for the pulsed emission
shows usually (in a favorable configuration) a double peak structure.
In this case, there exist a simple relation between this double
gamma-ray pulse separation, denoted by $\Delta$, and the geometry of
the system, assuming a rotating dipole with magnetic obliquity~$\chi$
and line of sight inclination~$\zeta$. Indeed, to understand the
physical mechanism, let us assume that the wind radiates mostly when
crossing a spherical shell located at a radius~$R_0$. Due to the
strong relativistic beaming effect for ultra relativistic expansion,
the observer only sees light travelling towards him, i.e. in the
direction $(\vartheta=\zeta, \varphi=\pi/2)$. For particles located in
the current sheets, this corresponds to an observer time$~t_{\rm s}$
satisfying
\begin{eqnarray}
  \label{eq:TS}
  R_0 & = & r_{\rm s}(\vartheta,\varphi,t_{\rm s}) \\
  & = & \beta_{\rm v} \,
  \rlight \left[ \pm \arccos (-\cot \zeta \, \cot \chi) +
    \frac{c\,t_{\rm s}}{\rlight} - \frac{\pi}{2} + 2 \, l \, \pi \right] \nonumber
\end{eqnarray}
where $l\in\mathbb{Z}$ is a natural integer. Solving for this
time~$t_s$ at which the photon leaves the current sheet, we find
\begin{equation}
  \label{eq:TS2}
  t_{\rm s}^\pm = \frac{R_0}{c\,\beta_{\rm v}} \pm \frac{\arccos (-\cot
    \zeta \, \cot \chi)}{\Omega_*} + \frac{P}{4} - l \, P
\end{equation}
The $\pm$ sign labels the arm of the double spiral structure (seen in
the equatorial plane) responsible for the radiation. Recall that the
double peak light-curves follow from this double spiral shape. The
time separation between two consecutive pulses becomes therefore
\begin{equation}
  \label{eq:SeparationPic}
  \Delta t_{\rm cons} = t_{\rm s}^+ - t_{\rm s}^- = 2 \, \frac{\arccos (-\cot
    \zeta \, \cot \chi)}{\Omega_*}
\end{equation}
Note that this is not necessarily the high-energy peak separation
because it can in principle vary between zero and a full period~$P$.
To be more precise, we have to compare one specified pulse with its
two neighbors in time, the one coming earlier to the observer and the
other later.  Doing so we get
\begin{eqnarray}
  \label{eq:SeparationPic2}
  \Delta t_{\rm peak} & = & {\rm min}(|t_{\rm s}^+ - t_{\rm s}^-|, |t_{\rm
    s}^+ - ( t_{\rm s}^- + P )| ) \\
  & = & {\rm min}(|\Delta t_{\rm cons}|, |\Delta
  t_{\rm cons} - P| )
\end{eqnarray}
Normalizing to the period of the pulsar~$P$, the phase separation
reads
\begin{eqnarray}
  \label{eq:SeparationPic3}
  \Delta & = & \frac{\Delta t_{\rm peak}}{P} =  {\rm min}\left( \frac{\arccos (-\cot
      \zeta \, \cot \chi)}{\pi}, \right. \\
  & & \left. \left| \frac{\arccos (-\cot
        \zeta \, \cot \chi)}{\pi} - 1 \right|\right)
\end{eqnarray}
From the definition of the arccos function, we get
\begin{equation}
  \label{eq:SeparationPic4}
  \Delta = 
  \begin{cases}
    \frac{\arccos (-\cot \zeta \, \cot \chi)}{\pi} \textrm{ if
  } \cot \zeta \, \cot \chi \le 0 \\
  1 - \frac{\arccos (-\cot \zeta \, \cot \chi)}{\pi} \textrm{ if
  } \cot \zeta \, \cot \chi \ge 0
  \end{cases}
\end{equation}
Finally, the relation between obliquity~$\chi$, line of sight
inclination angle~$\zeta$ and peak separation~$\Delta$ can be rearranged
into
\begin{equation}
  \label{eq:SeparationPic5}
  \cos(\pi\,\Delta) = 
  \begin{cases}
    - \cot \zeta \, \cot \chi \textrm{ if
  } \cot \zeta \, \cot \chi \le 0 \\
  \cot \zeta \, \cot \chi \textrm{ if
  } \cot \zeta \, \cot \chi \ge 0
  \end{cases}
\end{equation}
To summarize we find
\begin{equation}
  \label{eq:SeparationPic6}
  \cos(\pi\,\Delta) = |\cot \zeta \, \cot \chi|
\end{equation}
This relation is shown in Fig.~\ref{fig:SeparationPic} where $\cos
\zeta$ is plotted versus $\cos \chi$ for constant value of the
separation $\Delta$ starting from zero to $0.5$ with a step
$0.02$.  

A few special cases are worth to mention. First, zero separation or
overlapping of both pulses, corresponding to~$\Delta=0$, implies $\zeta
= \pi/2-\chi$.  For the chosen variables in
Fig.~\ref{fig:SeparationPic}, it corresponds to the quarter circle of
radius unity centered at the origin. Second, a separation of half a
period $\Delta=0.5$, implies a line of sight contained in the
equatorial plane of the pulsar, $\zeta = \pi/2$. It is independent of
the obliquity. This corresponds to the straight horizontal line along
the x-axis in Fig.~\ref{fig:SeparationPic}.
The relation Eq.~(\ref{eq:SeparationPic6}) was already noticed by
\cite{2005MmSAI..76..494K}, see figure therein.  For symmetry reasons,
we will assume that $\chi \in [0,\pi/2]$, therefore $\cot\chi\ge0$.

\subsection{The radio lag~$\delta$-peak separation~$\Delta$ relation}

Next, we show that the combination polar cap/striped wind allows to
derive a simple analytical relation between the lag of radio vs
gamma-ray arrival time~$\delta$ and the high-energy peak
separation~$\Delta$.  Because the two sites of emission are very
distinct, inside and outside the light cylinder, the lag is
interpreted as a time of flight delay between the region of radio
radiation and the one for gamma-ray photon production.

The choice of origin of time and line of sight located in the
plane~$(Oyz)$ (therefore $\varphi = \frac{\pi}{2}$) implies that the
radio photon emitted towards the observer will happen at periodic
dates with period $P$, for one pole, given by
\begin{equation}
  \label{eq:EmissionRadioNord}
  t_{\rm pc}^{\rm n} = \frac{\pi}{2\,\Omega_*} + k \, \frac{2\,\pi}{\Omega_*}
  = \frac{P}{4} + k \, P
\end{equation}
where $k\in\mathbb{Z}$ is a natural integer. The superscript~n denotes
bundles of photons emanating from the same polar cap, let us say the
north magnetic pole.  The same applies to the south cap but due to the
symmetry, the phase difference corresponds to half a period, $P/2$,
thus
\begin{equation}
  \label{eq:EmissionRadioSud}
  t_{\rm pc}^{\rm s} = t_{\rm pc}^{\rm n} + \frac{P}{2} = \frac{3\,P}{4} + k \, P
\end{equation}
The observer remains at a distance~$D$ from the center of the neutron
star.  The polar cap photon needs therefore a time $\Delta t = ( D -
R_* ) / c$ to reach him after emission from the pulsar. Thus the
detection time is the sum
\begin{equation}
  \label{eq:TempsArrivee1}
  t_1 = t_{\rm pc}^{\rm n} + \frac{D-R_*}{c}
\end{equation}
The gamma-ray photon from the current sheets arrives earlier because
closer to the observer, he needs a shorter flying time given by
$\Delta t=(D-R_0)/c$ leading to a time of detection given by
\begin{equation}
  \label{eq:TempsArrivee2}
  t_2 = t_{\rm s}^\pm + \frac{D-R_0}{c}
\end{equation}
As a consequence, the difference in arrival time between both kind of
photons, focusing on the radio emission coming out of the north pole
(visible for $0<\zeta<\pi/2$), is
\begin{eqnarray}
  \label{eq:TempsArrivee3}
  t_a & = & t_2 - t_1 = t_{\rm s}^\pm - t_{\rm pc}^{\rm n} + \frac{R_*-R_0}{c} \\
  & = & \frac{1-\beta_{\rm v}}{\beta_{\rm v}} \, \frac{R_0}{c} \pm \frac{\arccos (-\cot
    \zeta \, \cot \chi)}{\Omega_*} + \frac{R_*}{c} - ( k + l ) \, P \nonumber
\end{eqnarray}
By normalizing this time to the period of the pulsar~$P$ and using
Eq.~(\ref{eq:SeparationPic4}) with $\cot \zeta \, \cot \chi \ge 0$, we
arrive at
\begin{eqnarray}
  \delta & = & \frac{t_a}{P} \\
  \label{eq:TempsArrivee5}
  & = & \frac{1-\beta_{\rm v}}{\beta_{\rm v}} \, \frac{R_0}{2\,\pi\,\rlight} \pm \frac{1-\Delta}{2} + \frac{R_*}{2\,\pi\,\rlight} - ( k + l )
\end{eqnarray}
The same procedure applied to the south pole with $\pi/2<\zeta<\pi$
implies $\cot \zeta \, \cot \chi \le 0$, leading to the same
conclusion because of symmetry.

\begin{figure}
  \centering
  \includegraphics[width=0.45\textwidth]{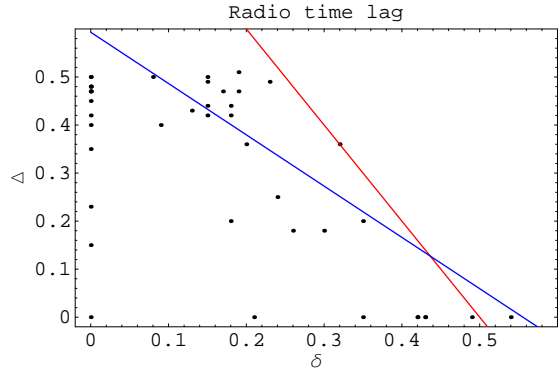}
  \caption{Gamma-ray peak separation~$\Delta$ vs radio time
    lag~$\delta$ as observed by Fermi-LAT. The blue curve is a linear
    fit and the red straight line the upper limit as predicted by
    Eq.~(\ref{eq:deltaD}).}
  \label{fig:Delai}
\end{figure}

Recall that in order to observe pulsed emission from the radially
expanding wind, the flow has to be relativistic with $\Gamma_{\rm v}
\gg 1$ and $R_0$ must lie close enough to the light-cylinder such that
the underlying condition is fulfilled
\begin{equation}
  \label{eq:Pulsation}
  R_0 \lesssim 2\,\pi\,\Gamma_{\rm v}^2\,\rlight
\end{equation}
Thus the upper limit to the first term in
Eq.~(\ref{eq:TempsArrivee5}) is
\begin{equation}
  \label{eq:Cste}
  \frac{1-\beta_{\rm v}}{\beta_{\rm v}} \, \frac{R_0}{2\,\pi\,\rlight}
  \approx 
  \frac{1}{2\,\Gamma_{\rm v}^2} \, \frac{R_0}{2\,\pi\,\rlight}
  \lesssim \frac{1}{2}
\end{equation}
Moreover, for all the pulsars in the catalog, even for the millisecond
pulsars, the light-cylinder radius is much larger than the stellar
radius, $\rlight\gg R_*$. To a good approximation, we can neglect the
term $\frac{R_*}{2\,\pi\,\rlight}$ in
Eq.~(\ref{eq:TempsArrivee5}), the neutron star
is located well within its light-cylinder. The remaining expression
for a positive lag is
\begin{equation}
  \label{eq:deltaD}
  \delta \approx \frac{1-\Delta}{2}
\end{equation}
This has to be compared with the $\delta-\Delta$ diagram published in
\cite{2010ApJS..187..460A}. The data points as well as the relation
Eq.~(\ref{eq:deltaD}) are shown in Fig.~\ref{fig:Delai}.  All the
points but one lie below the line given by Eq.~(\ref{eq:deltaD}).  The
spread in radio lag along this line can be understood as a fluctuation
in the precise location of the most significant emission in the wind.

Finally, we give an expression for the linear fit to the sample of
pulsars showing double peaked gamma-ray light-curves simultaneously
with radio emission by
\begin{equation}
  \label{eq:Fit}
  \Delta_{\rm fit} \approx 0.59 - 1.06 \, \delta
\end{equation}
shown in blue line in Fig.~\ref{fig:Delai}.

\subsection{Other retardation effects}

Including general-relativistic effects do not change much the time lag
between radio and gamma-ray photons. Indeed using the Schwarzschild
geometry, for a photon with radial motion, the time required to reach a
distant observer located at a distance $D$, starting on the stellar
surface is
\begin{equation}
  \label{eq:RetardGR1}
  c \, \Delta t_{\rm pc} = D - R_* + R_s \, \ln \left( \frac{D-R_s}{R_*-R_s} \right) 
\end{equation}
$R_s = 2\,G\,M/c^2$ being the Schwarzschild radius. A similar expression
applies for gamma-rays
\begin{equation}
  \label{eq:RetardGR2}
  c \, \Delta t_{\gamma} = D - R_0 + R_s \, \ln \left( \frac{D-R_s}{R_0-R_s} \right) 
\end{equation}
Therefore the time of flight difference compared to Newtonian gravity
is 
\begin{equation}
  \label{eq:RetardGR3}
  \Delta t_{\rm GR} = \Delta t_{\gamma} - \Delta t_{\rm pc} =
  \frac{R_* - R_0}{c} + \frac{R_s}{c} \, \ln \left( \frac{R_*-R_s}{R_0-R_s} \right)
\end{equation}
Normalized to the pulsar period we get an increase by an amount of
\begin{equation}
  \label{eq:RetardGR4}
  \delta_{\rm GR} = \frac{R_s}{2\,\pi\,\rlight} \, \ln \left( \frac{R_*-R_s}{R_0-R_s} \right)
\end{equation}
For neutron star parameters, this remains negligible compared
to unity because $R_s\ll\rlight$. Typical values are $\delta_{\rm GR}
\approx 0.01$.

\subsection{Phase-inclination diagram}

\begin{figure*}
  \centering
  \begin{tabular}{cc}
    \includegraphics[width=0.45\textwidth]{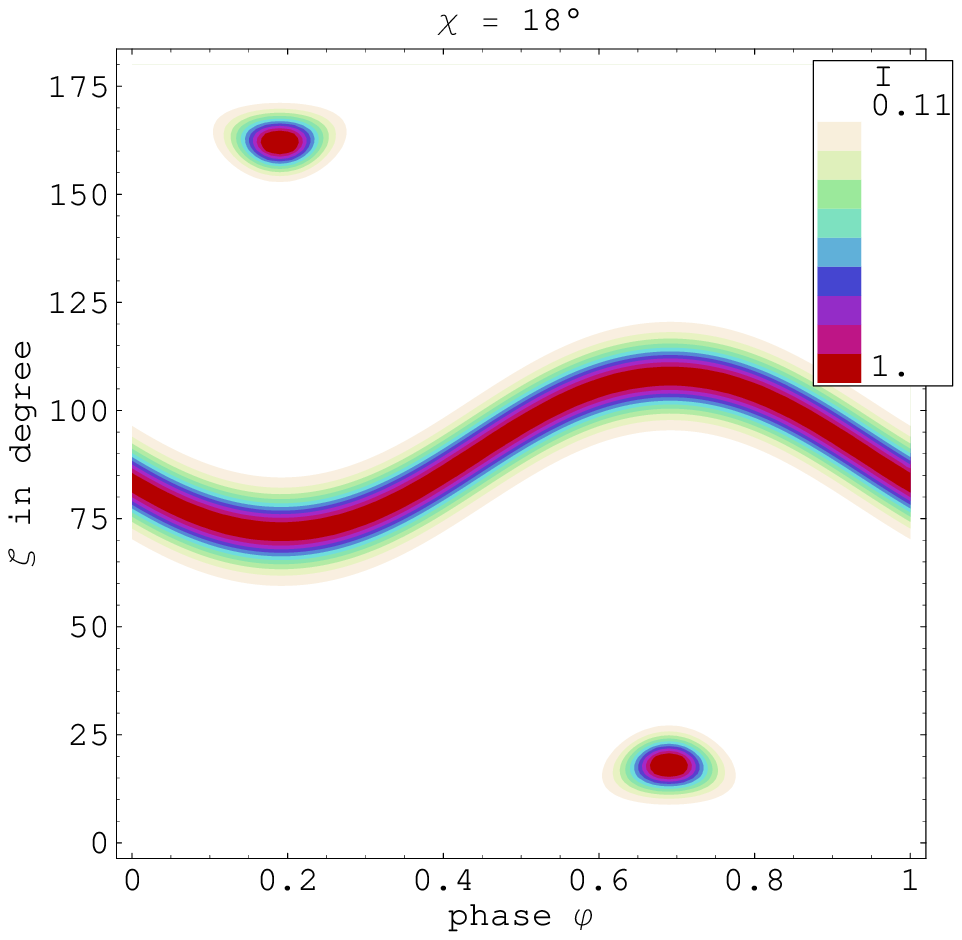} &
    \includegraphics[width=0.45\textwidth]{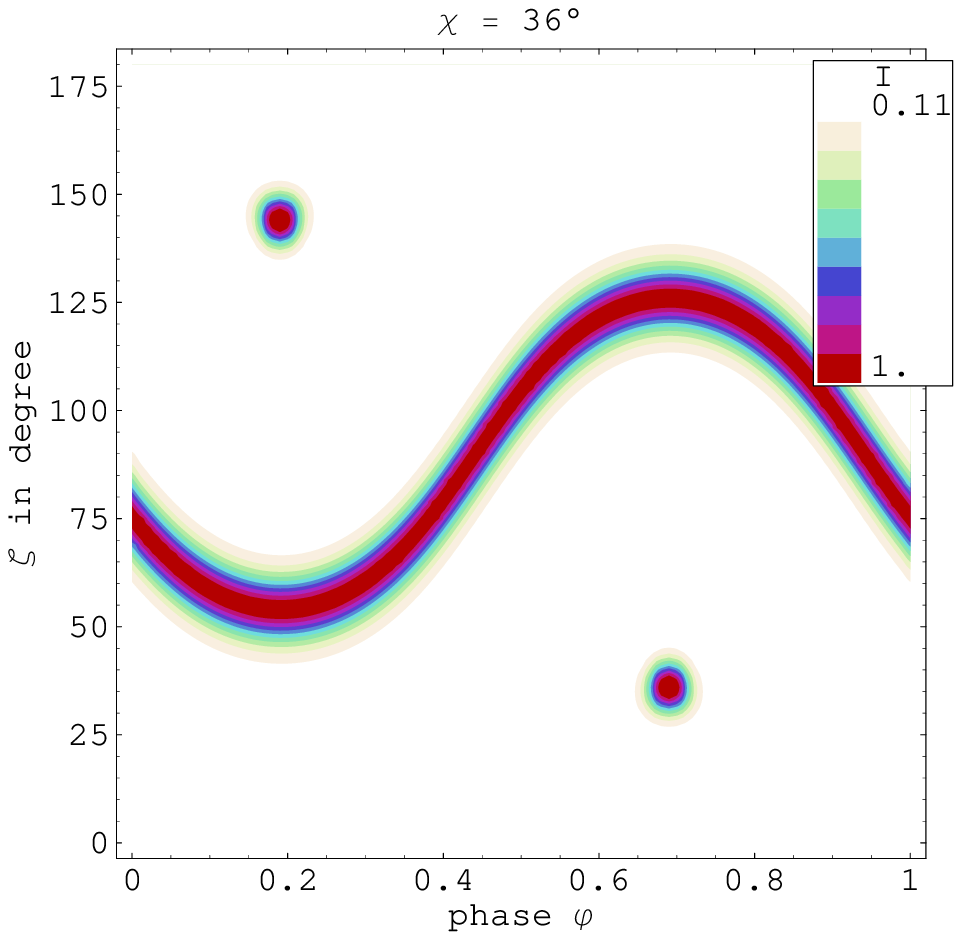} \\
    \includegraphics[width=0.45\textwidth]{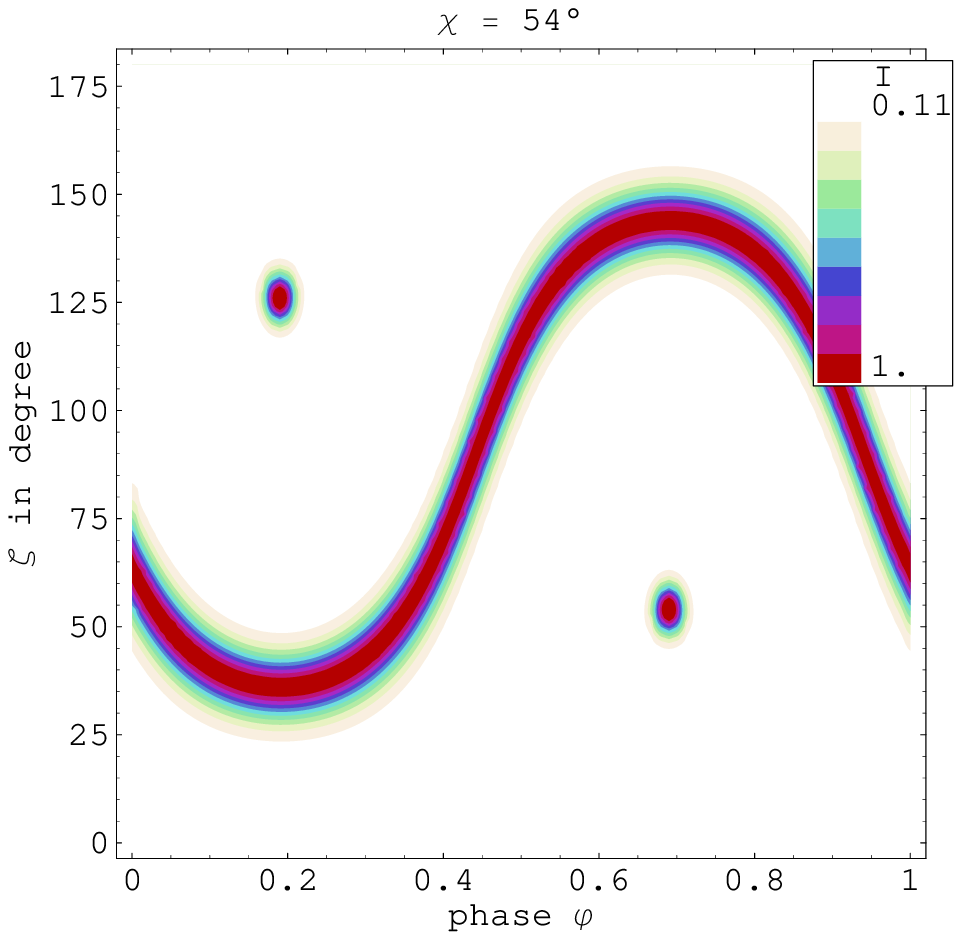} &
    \includegraphics[width=0.45\textwidth]{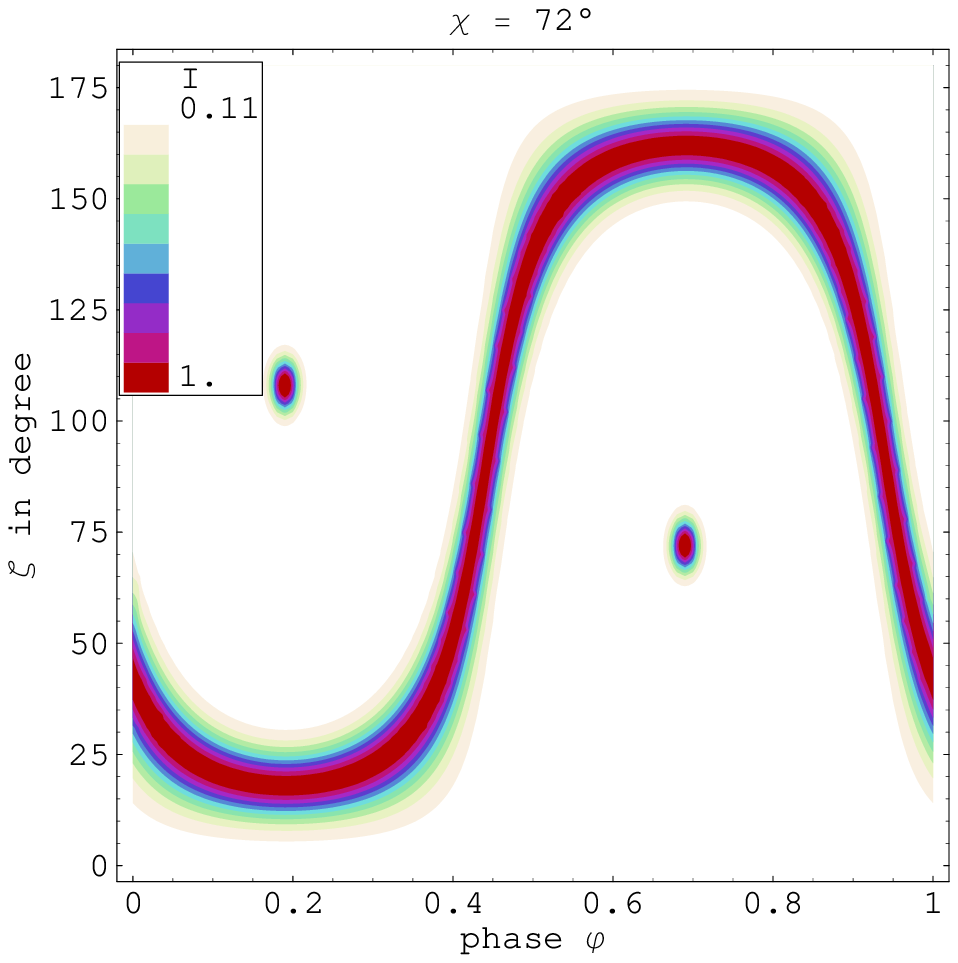}
  \end{tabular}  
  \caption{Phase plot of the pulsed gamma-ray and radio emission
    components for a full period of the pulsar (phase $\in[0,1]$) for
    an inclination of the line of sight~$\zeta$ between 0 and 180
    degrees and magnetic obliquities $\chi=18,36,54$ and~$72$ degrees,
    from top left to bottom right. Note the colour-coded range shown
    in the legends of each phase plot: white colour corresponds to the
    faintest phase (off pulse) with intensity around 0.11 whereas red
    colour corresponds to the brightest luminosity i.e. 1 (in
    normalized unities).}
  \label{fig:PhaseInclination}
\end{figure*}

Next, we show the phase-inclination diagram for different inclinations
of the line of sight~$\zeta$ as well as different magnetic
obliquities~$\chi$ of the pulsar. This is done for the high-energy
photons as well as for the radio emission coming from the polar cap,
see Fig.~\ref{fig:PhaseInclination}. In order to point out the
relative phase shift between both emission processes, we normalized
independently their respective maximum intensity to unity. The
continuous rotated S-shaped structure is a characteristic of the
striped wind whereas the two spots correspond to a mapping of the two
polar caps. Both structures are symmetric with respect to the
equatorial plane, i.e. compared to an inclination $\zeta=90^o$, as
expected from the geometry.

The observer will see different kind of light-curves depending on his
viewing angle. First, radio photons are observable only when the line
of sight intersects at least one polar cap, in the vicinity of the
magnetic poles, that is when $\zeta \approx \chi \pm \vartheta_{\rm
  pc}$. Second, to observe the gamma-ray pulsation, this same line of
sight should intersect the current sheets, or in other words
$\pi/2-\chi \le \zeta \le \pi/2+\chi$. Conversely, no gamma-ray pulses
are seen when $\zeta \le \pi/2-\chi$ or $\zeta \ge \pi/2+\chi$. Thus a
necessary condition to detect simultaneously radio and gamma-ray
photons is $\chi\ge\pi/4$ and $\zeta \approx \chi$. For those
particular pulsars, possessing a double gamma-ray peak and radio
emission, we are able to find the geometrical parameters from the peak
separation, see Fig.~\ref{fig:SeparationPic}.  Third, a single pulse
in the high-energy light-curve occurs when the line of sight passes
just through the edge of the striped part of the wind, this can be
interpreted as the special case $\zeta\approx\pi/2-\chi$.  In the very
special case of one gamma-ray pulse and observable radio emission,
this leads immediately to the geometrical parameters $\zeta \approx
\chi \approx \pi/4 = 45^o$. This seems to be the case for
PSRJ0437-4715 and PSRJ2229+6114.

\begin{figure*}
  \centering
  \begin{tabular}{cc}
    \includegraphics[width=0.45\textwidth]{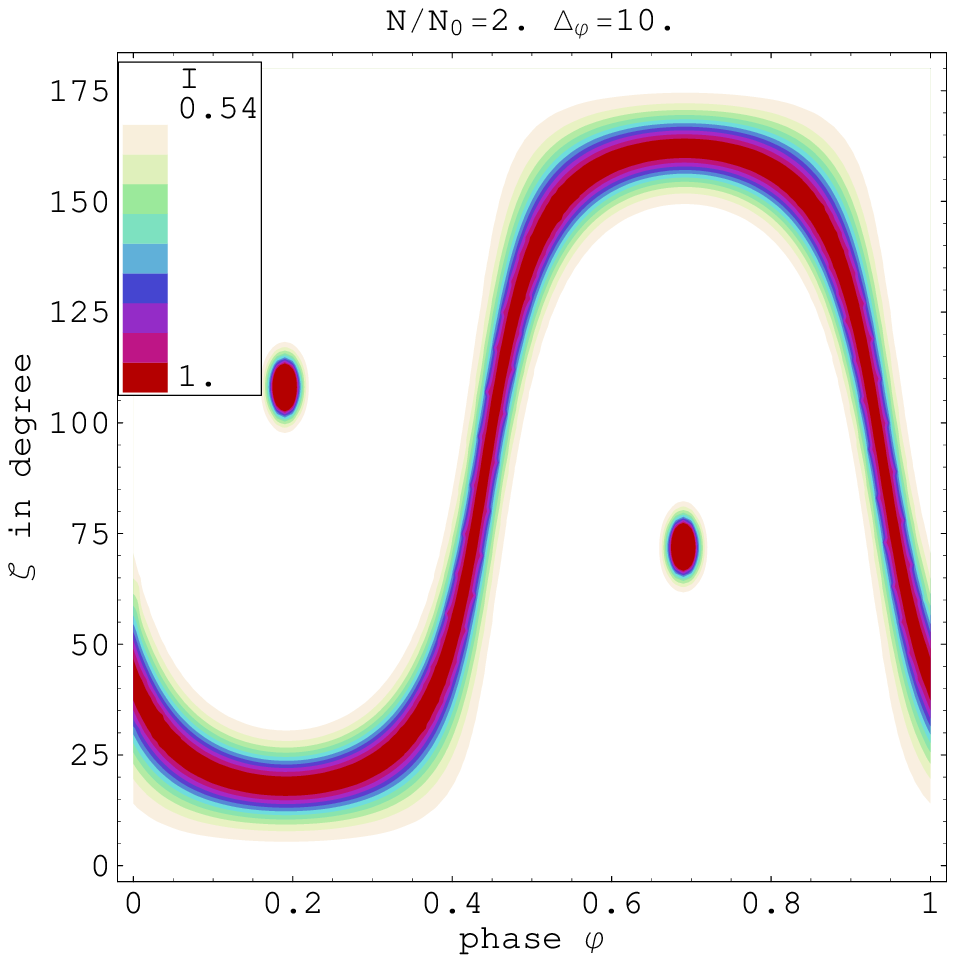} &
    \includegraphics[width=0.45\textwidth]{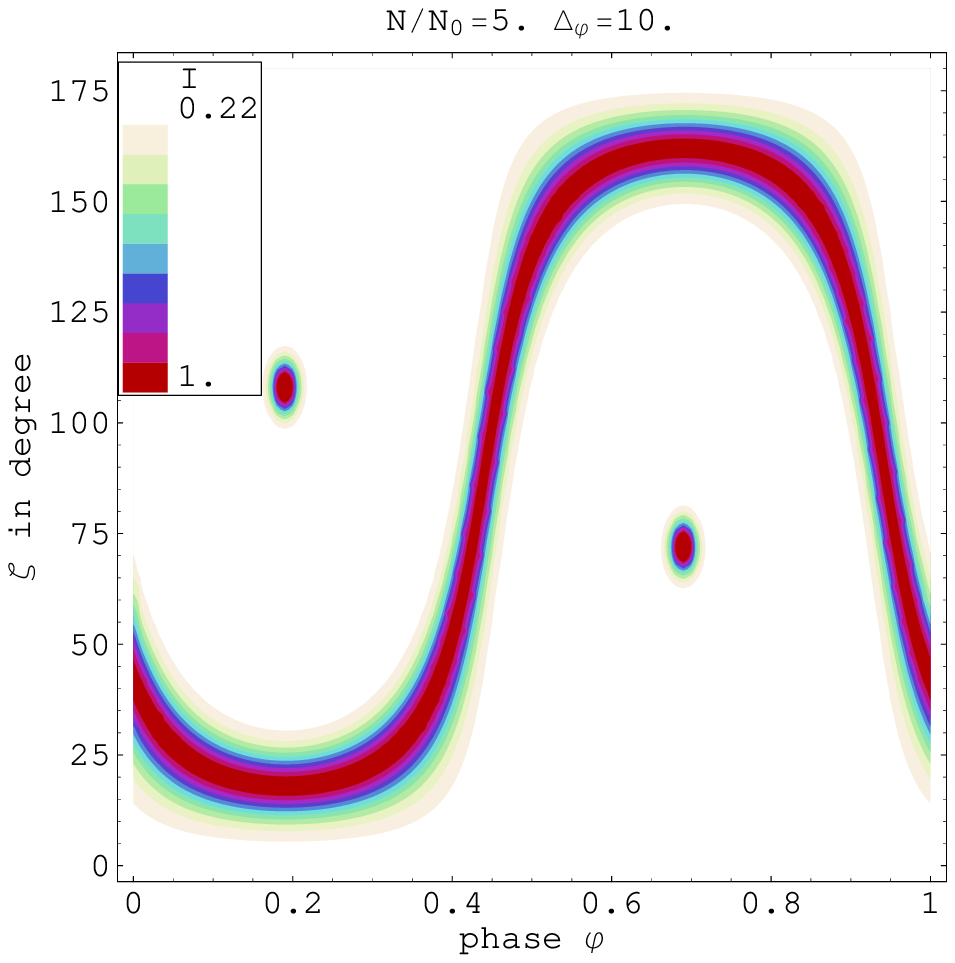}
  \end{tabular}  
  \caption{Phase plot of the pulsed gamma-ray and radio emission
    components for a full period of the pulsar (phase $\in[0,1]$) for
    an inclination of the line of sight~$\zeta$ between 0 and 180
    degrees and a magnetic obliquity $\chi=72$ degrees. Only the ratio
    $N/N_0$ changes.}
  \label{fig:PhaseInclinationN}
\end{figure*}

The phase plots shown in Fig.~\ref{fig:PhaseInclination} have been
computed for different inclination angles~$\zeta$ and
obliquities~$\chi$ but with constant current sheet thickness,
parameterized by $\Delta_\varphi$, and constant particle density
contrast, which means constant ratio $N/N_0$. More precisely, we
adopted $\Delta_\varphi = 10$ and $N/N_0=10$. This last value explain
the factor around~10 in intensity between off-pulse and on-pulse
phase.  Indeed, by decreasing the fluctuation in particle density
between the hot unmagnetized and cold magnetized part of the wind, as
done in Fig.~\ref{fig:PhaseInclinationN}, we recognize a similar trend
in the intensity diagram. If the ratio $N/N_0=5$, the luminosity
varies between $0.22\approx1/5$ and $1$ in normalized unities whereas
for $N/N_0=2$, it lies between $0.54\approx1/2$ and 1.

\begin{figure*}
  \centering
  \begin{tabular}{cc}
    \includegraphics[width=0.45\textwidth]{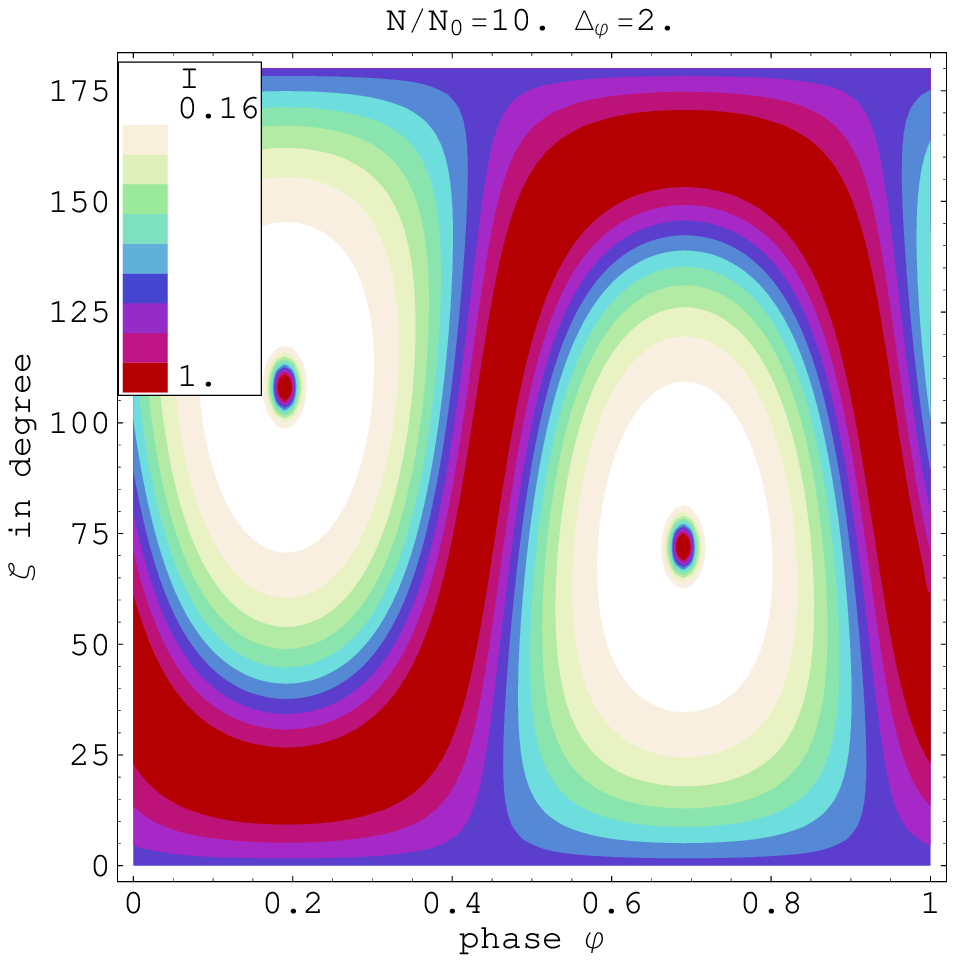} &
    \includegraphics[width=0.45\textwidth]{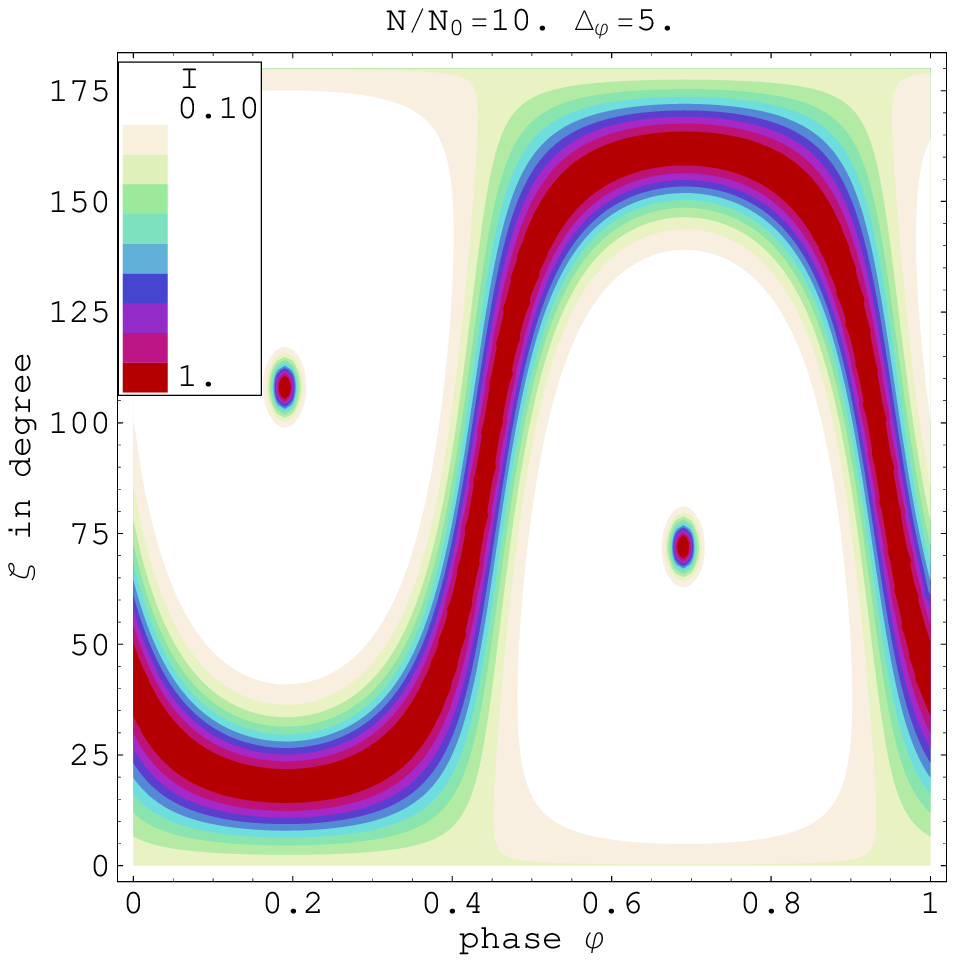}
  \end{tabular}  
  \caption{Phase plot of the pulsed gamma-ray and radio emission
    components for a full period of the pulsar (phase $\in[0,1]$) for
    an inclination of the line of sight~$\zeta$ between 0 and 180
    degrees and a magnetic obliquity $\chi=72$~degrees. Only the
    parameter $\Delta_\varphi$ varies.}
  \label{fig:PhaseInclinationD}
\end{figure*}

Moreover, the current sheet thickness directly impacts on the duty
cycle of the light-curve, or in other words, on the width of the
gamma-ray pulses. This has been check by changing the parameter
$\Delta_\varphi$ to 5 or 2 instead of the previous value of 10.
Results are shown in Fig.~\ref{fig:PhaseInclinationD}.

\begin{figure*}
  \centering
  \begin{tabular}{cc}
    \includegraphics[width=0.45\textwidth]{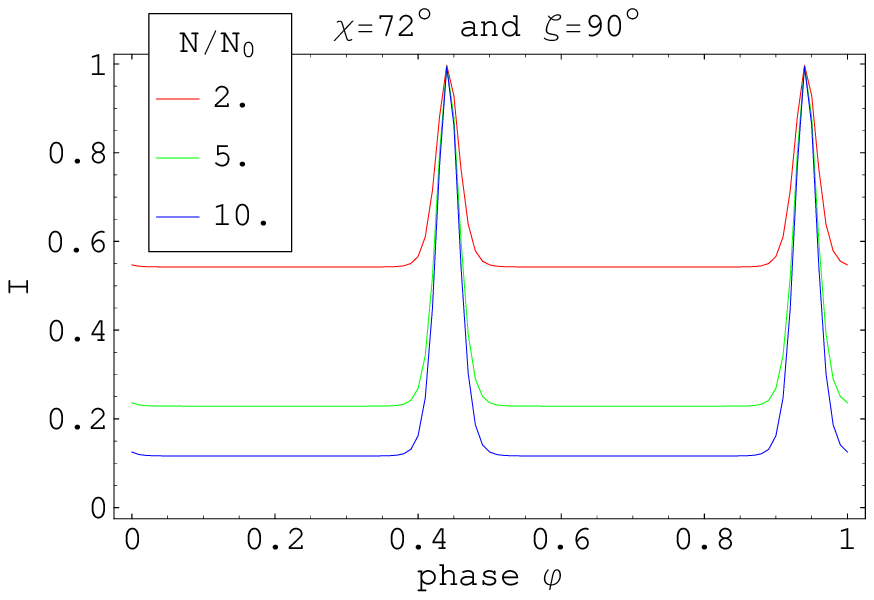} &
    \includegraphics[width=0.45\textwidth]{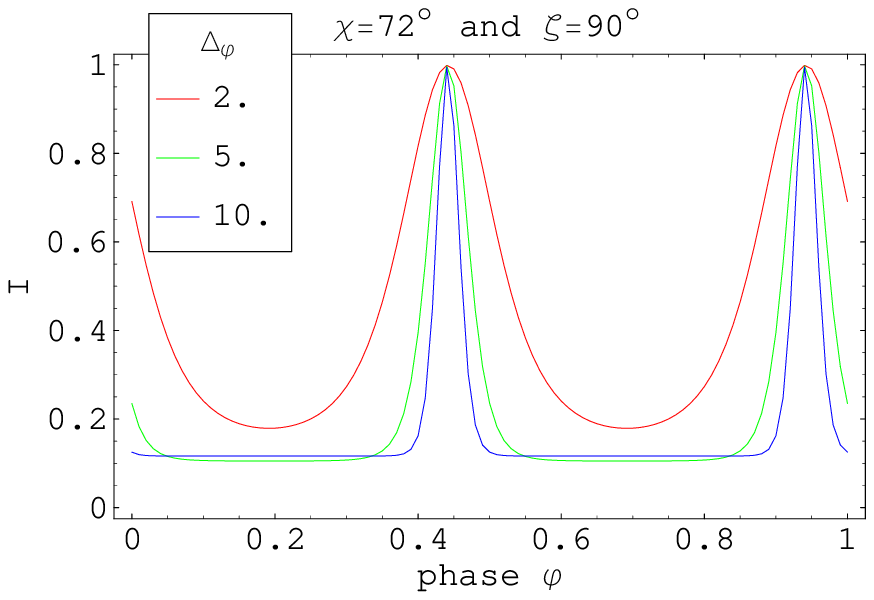}
  \end{tabular}  
  \caption{Light-curves of the pulsed gamma-ray emission component,
    for a full period of the pulsar (phase $\in[0,1]$) and for an
    inclination of the line of sight~$\zeta=90$~degrees and a magnetic
    obliquity $\chi=72$~degrees. On the left picture, for different
    parameters $N/N_0$ with $\Delta_\varphi=10$ and on the right
    picture for different $\Delta_\varphi$ with $N/N_0=10$.}
  \label{fig:PhaseInclinationSummary}
\end{figure*}

To better assess the differences, we summarize all the light-curves
for $\chi=72$~degrees and $\zeta=90$~degrees in two plots as depicted
in Fig.\ref{fig:PhaseInclinationSummary}.


Finally, there exist two limiting cases. First the aligned
rotator showing no pulsed emission. Second the perpendicular rotator
emits pulsed high-energy radiation over the whole sky and both its
polar caps are visible. A direct consequence is a double peak
structure in both radio and gamma-ray light-curves, with a separation of
$\Delta=0.5$.  PSRJ0030+0451 is a typical example.

\subsection{Flux correction factor}

\begin{figure} 
  \centering
  \includegraphics[width=0.45\textwidth]{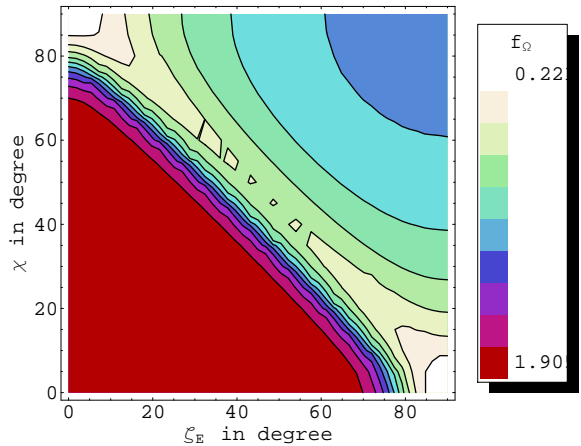}
  \caption{Flux correction factor $f_\Omega$ for the striped wind
    model vs obliquity $\chi$ and inclination of Earth line of sight
    $\zeta_E$. Two different regions with roughly constant factor are
    clearly distinguishable along $\zeta_E = \pi/2 - \chi$.}
  \label{fig:FacteurCorrection}
\end{figure}

The photon flux $F_{\rm obs}$ measured by an observer located on Earth
is biased due to anisotropic emission from the wind depending on the
viewing angle $\zeta$. This fact is clear from the aforementioned
phase-inclination plots 
shown in the previous paragraphs. The observed gamma-ray flux has to
be corrected to obtain the true gamma-ray luminosity by introducing a
correction factor~$f_\Omega$ defined by
\begin{equation}
  \label{eq:LuminositeGamma}
  L_\gamma = 4 \, \pi \, f_\Omega \, F_{\rm obs} \, D^2
\end{equation}
Here $D$ is the distance of the pulsar to the observer and $F_{\rm
  obs}$ the observed flux.  As in the polar cap and outer/slot gap
models \citep{2009ApJ...695.1289W}, the correction implied here by a
relativistic beaming effect is given by
\begin{equation}
  \label{eq:FacteurFocal}
  f_{\Omega}(\chi,\zeta_E) = \frac{\int_0^{\pi} \int_0^{2\,\pi}
    F_\gamma(\chi,\zeta,\varphi) \, \sin\zeta \, d\zeta \, d\varphi}
  {2 \, \int_0^{2\,\pi} F_\gamma(\chi,\zeta_E,\varphi) \, d\varphi}
\end{equation}
For the striped wind model, this correction factor is shown in
fig.~\ref{fig:FacteurCorrection} with the full dependence on obliquity
$\chi$ and inclination of Earth line of sight
$\zeta_E$.
We can approximately separate the correction factor into two regions
of constant value. In the first region, for an obliquity $\chi<\pi/2 -
\zeta_E$, the correction is close to be uniform and equal roughly to
1.90. This case corresponds to a line of sight not crossing the
current sheets in the wind, there is almost no pulsed emission
visible. In the second region, where $\chi > \pi/2 - \zeta_E$, the
correction is also almost uniform and equal roughly to 0.4. This case
corresponds to a line of sight intersecting the current sheets,
leading to pulsed emission. This behavior is expected from the
definition of the correction factor Eq.~(\ref{eq:FacteurFocal}).
Indeed, for a fixed obliquity $\chi$, the numerator is a constant
whereas the denominator depends on the line of sight towards the Earth. On
one side, in the first region $\chi > \pi/2 - \zeta_E$, the
phase-averaged emission is faint and weakly pulsed. It follows a small
denominator therefore a large correction factor. On the other side, in
the second region, the situation is opposite, the emission during the
pulses is strong and so the denominator of Eq.~(\ref{eq:FacteurFocal})
larger, finally the correction factor is weakest.

Note that these bounding values depend on the sheet thickness
parameterized by $\Delta_\varphi$ as well as on the particle density
contrast parameterized by $N_0$ and $N$.

\subsection{Light-curve fitting and gamma-ray luminosity}

We conclude our study by fitting light-curves of a small sample of
pulsars.  The only relevant free parameters in our model are the
geometry of the wind, the particle density number and the size of the
current sheets.  In this last section, we specialize our results to
some Fermi detected gamma-ray pulsars and show the best parameters
fitting their high-energy light-curves above 100~MeV.  Therefore, the
knowledge of the viewing angle and the obliquity allows an estimation
of the flux correction via the beaming factor. Eventually a true
gamma-ray luminosity versus spin-down luminosity can be plotted.

We start with an estimate of the peak separation when one radio pulse
is detected. In that case, $\zeta\approx\chi$ as explained above and
the knowledge of $\Delta$ immediately implies a solution for $\zeta$.
This has been done for several pulsars and listed in
Tab.~\ref{tab:Geometrie}. In all the results shown below, for
simplicity, we took a constant Lorentz factor of the wind equal to
$\Gamma_{\rm v}=10$.

\begin{table*}
  \centering
  \begin{tabular}{cccccccccccc}
    \hline
    Pulsar & $\delta$ (obs) & $\Delta$ (obs) & $\chi$ & $\zeta$ &
    $\delta$ (model) & $\Delta_\varphi$ & $ N/N_0$ & $f_\Omega$ &
    $L_\gamma$ & $L_\gamma$ (corrected) & efficiency $\eta$ \\
    & & & deg & deg & $(1-\Delta)/2$ & & & & $10^{26}$~W &
    $10^{26}$~W & \\
    \hline
    \hline
    J0030+0451       & 0.18 & 0.44 & 67 & 85 & 0.28  &  5 & 10 & 1.17
    & 0.57 & 0.49 & 0.16 \\
    J0218+4232       & 0.32 & 0.36 & 57 & 75 & 0.32  &  3 &  3 & 1.10
    & 27-69 & 24-62 & 0.10-0.26 \\
    J0437-4715       & 0.43 & -    & 45 & 40 & 0.5     & 10 & 10 & 1.07
    & 0.054 & 0.050 & 0.016 \\
    J1124-5916       & 0.23 & 0.49 & 80 & 85 & 0.255 &  5 &  3 & 1.20
    & 100 & 83 & 0.007 \\
    J2021+3651       & 0.17 & 0.47 & 73 & 65 & 0.265 & 10 & 10 & 1.14
    & 250 & 220 & 0.065 \\
    J2032+4127       & 0.15 & 0.50 & 89 & 70 & 0.25  &  5 &  3 & 1.18
    & 34-170 & 29-145 & 0.11-0.55 \\
    J2043+2740       & 0.20 & 0.36 & 57 & 65 & 0.32  &  5 &  3 & 1.05
    & 6 & 5.7 & 0.10 \\
    J2229+6114       & 0.49 & -    & 45 & 40 & 0.5     & 10 & 10 & 1.07
    & 17-1100 & 16-1026 & 0.0007-0.045 \\
    \hline
  \end{tabular}
  \caption{Geometry of the pulsar wind for 8 gamma-ray pulsars. The
    angle of inclination and obliquity are given in degrees. The
    Lorentz factor of the wind is the same for the whole sample and
    taken to be $\Gamma_{\rm v}=10$.}
  \label{tab:Geometrie}
\end{table*}

Let us have a deeper look on a representative sample of Fermi-detected
gamma-ray pulsars.

\subsubsection{PSR J0030+0451}

PSR J0030+0451 is a millisecond pulsar, $P=4.87$~ms, showing a double
pulse structure in the radio band. This may suggest that both its
magnetic poles are visible, or in other words, it is almost a
perpendicular rotator with $\chi \approx 90^o$. Nevertheless, the
maximal intensity of both radio pulses are sensibly different, this is
interpreted as the line of sight passing closer to one magnetic pole
than to the other.  An alternative would be to explain it by a process
occurring in the vicinity of the polar caps with variable efficiency.
However, polar cap emission is not the main purpose of this work so we
simply assume identical shapes for both radio-pulses.  Therefore, an
obliquity close to $90$~degrees but less matches the right geometry.
Setting $\chi \approx 67^o$ satisfactorily agrees with the radio
light-curve, see Fig.~\ref{fig:FermiCL}, on the top left plot.  To
have the right gamma-ray peak separation, we have to adopt $\zeta
\approx 85^o$.  Moreover, the radio-pulses are very broad, each of
them having a duty cycle of roughly $0.2$ in phase.  Therefore, to
reconcile our model with data, we have to extend the polar cap region
to a sizeable fraction of the whole neutron star surface,
Fig.~\ref{fig:FermiCL}, top left plot.  Emission starts right after
the light-cylinder radius. There is still a excess of 0.1 in the phase
delay compared to observation.  This has to be explained by some other
retardation effects of the radio pulse such as the strong
gravitational field regime (which we have shown to be negligible) or
by magnetic field bending due to charges flowing within the
magnetosphere and disturbing the closed field lines structure taking
to be an exact dipole.

\subsubsection{PSR J0218+4232}

PSR~J0218+4232 is another millisecond pulsar, $P=2.32$~ms, showing a
less clear double pulse structure in gamma-rays. Its radio-pulse shape
looks much more complicated with something like conal and core
component, over a wide range of the pulsar period. The best fit is
obtain for an obliquity $\chi=57^o$ and an inclination of
$\zeta=75^o$. The gamma-ray light curve adjusts well to Fermi data.
Moreover the gamma-ray pulse time lag compared to the middle of the
radio pulse matches precisely the measurements. Note that the
gamma-ray off pulse emission remains at an appreciable level over the
full period, see Fig.~\ref{fig:FermiCL}, on the top right plot.

\subsubsection{PSR J0437-4715}

PSR~J0437-4715 is a special case of millisecond pulsar, $P=5.76$~ms,
showing only one gamma-ray pulse combined with a sharp radio pulse. As
explained in a previous discussion, the unique geometry allowing such
a behavior needs an obliquity almost equal to the inclination of the
line of sight, namely $\zeta \approx \chi \approx 45^o$. Indeed, we
take $\chi=45^o$ and $\zeta=40^o$ and arrive at the light curve
presented in Fig.~\ref{fig:FermiCL}, second row, left plot. For one
gamma-ray pulse, our model predicts a delay of $0.5$ in phase between
radio and gamma-ray pulsars, in relative good agreement with Fermi
measuring $0.43$.  The overestimate is about 0.07.

\subsubsection{PSR J1124-5916}

PSR J1124-5916 shows a sharp double gamma-ray peak with significant
off-pulse emission in combination with a single radio peak. Adopting
the parameters $\chi=80^o$ and $\zeta=85^o$, our model is compared
with observation in Fig.~\ref{fig:FermiCL}, second row, right plot.
The expected time lag $0.255$ is not very different from the
measured~$0.23$.

\subsubsection{PSR J2021+3651}

PSR~J2021+3651 gamma-ray light-curve shows a narrow double pulse
structure and a single radio pulse. The peak separation is close to
$0.5$. Here again, for the best fit parameters $\chi=73^o$ and
$\zeta=65^o$, the predicted lag of $0.265$ overshoots the observation
by roughly $0.1$, Fig.~\ref{fig:FermiCL}, third row, left plot.

\subsubsection{PSR J2032+4127}

PSR J2032+4127 is an example of half-a-period peak separation implying
a line of sight inclination of $\zeta\lesssim90^o$. Moreover, because
we see radio pulse it should be almost a perpendicular rotator.
However, this would permit to detect both radio component which is not
the case, so we must conclude that the inclination is slightly
different from an orthogonal rotator with a small polar cap
reproducing a width of 0.1 in the radio pulse, Fig.~\ref{fig:FermiCL},
third row, right plot.  We found best agreement for $\chi=89^o$ and
$\zeta=70^o$.

\subsubsection{PSR J2043+2740}

It seems that PSR J2043+2740 shows evidence for three gamma ray
pulses, Fig.~\ref{fig:FermiCL}, bottom left plot. The two extreme ones
are significant whereas the middle one is less significant. We tried
to fit the light curve according to the two well separated pulses. The
time delay is overestimated by $0.12$.

\subsubsection{PSR J2229+6114}

\begin{figure*}
  \centering
  \begin{tabular}{cc}
  \includegraphics[width=0.45\textwidth]{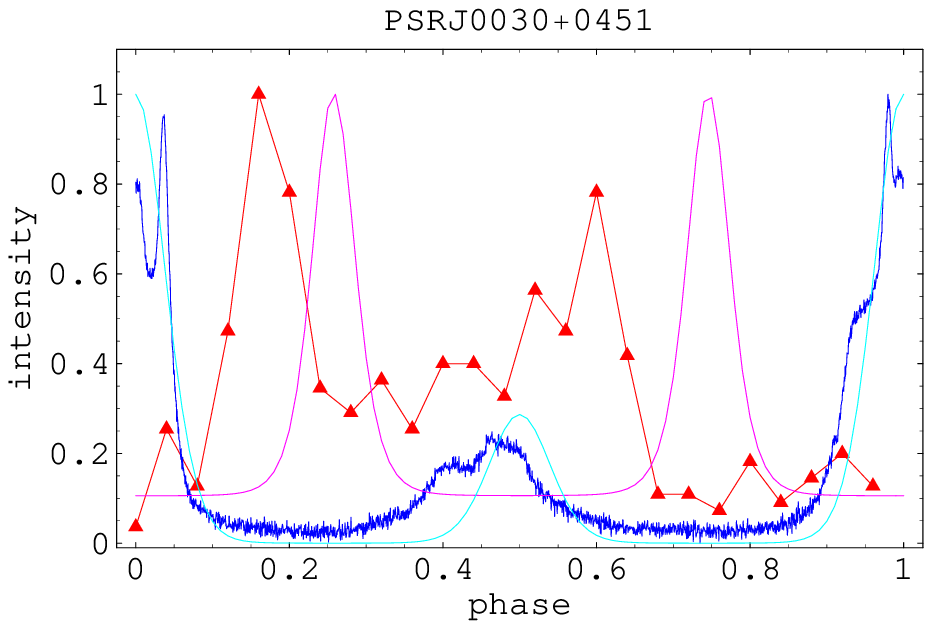} &
  \includegraphics[width=0.45\textwidth]{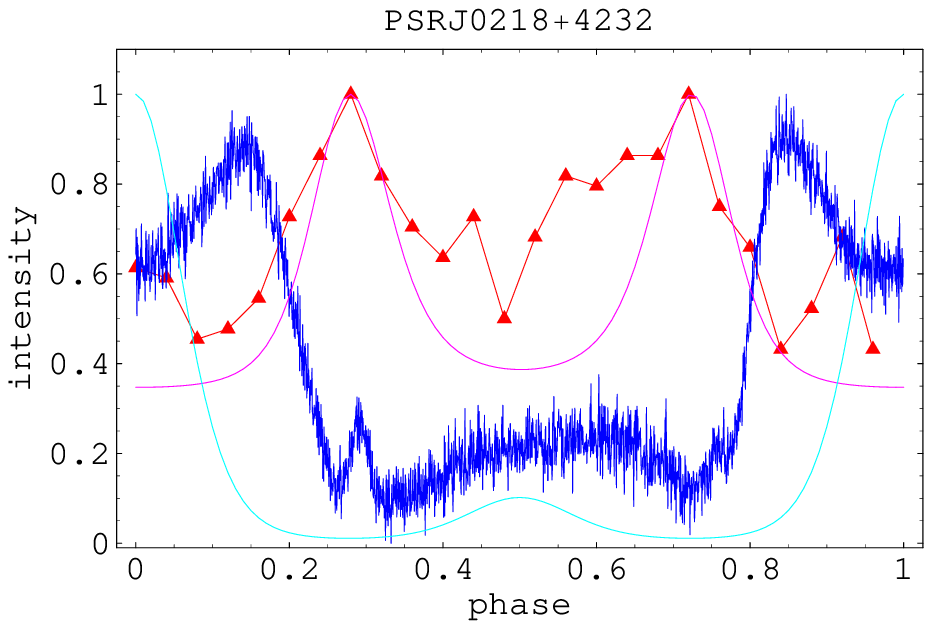} \\  
  \includegraphics[width=0.45\textwidth]{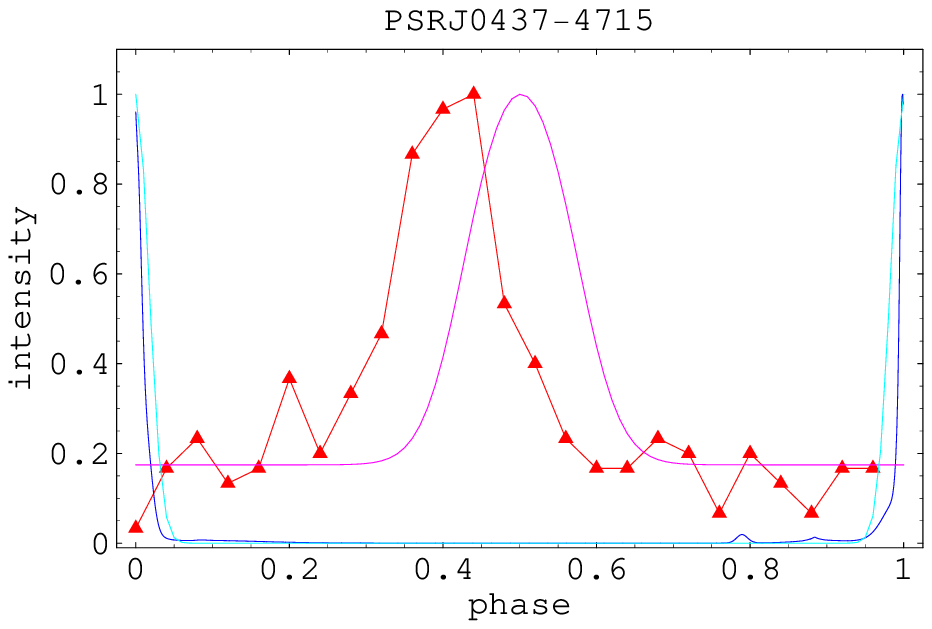} &
  \includegraphics[width=0.45\textwidth]{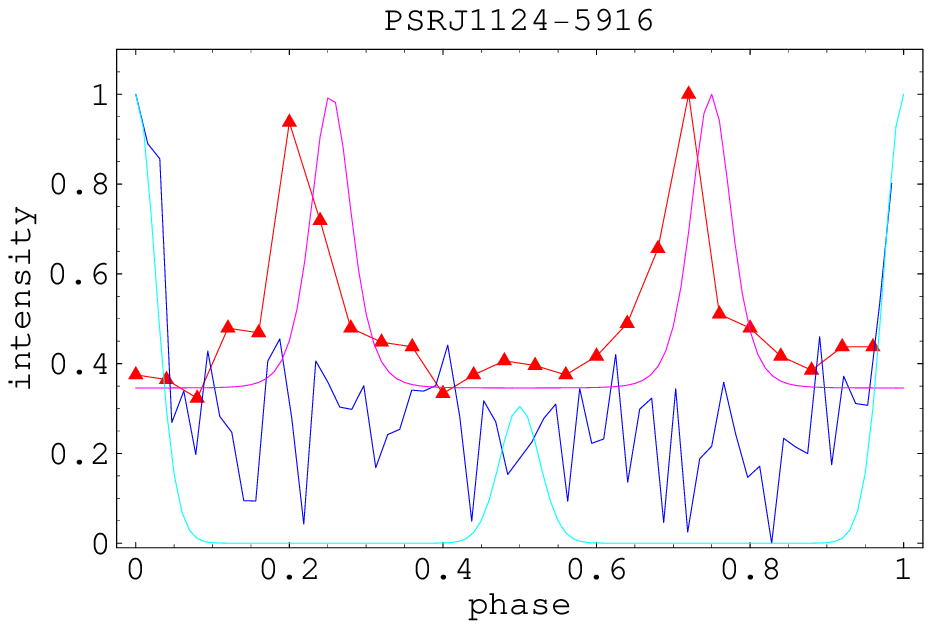} \\
  \includegraphics[width=0.45\textwidth]{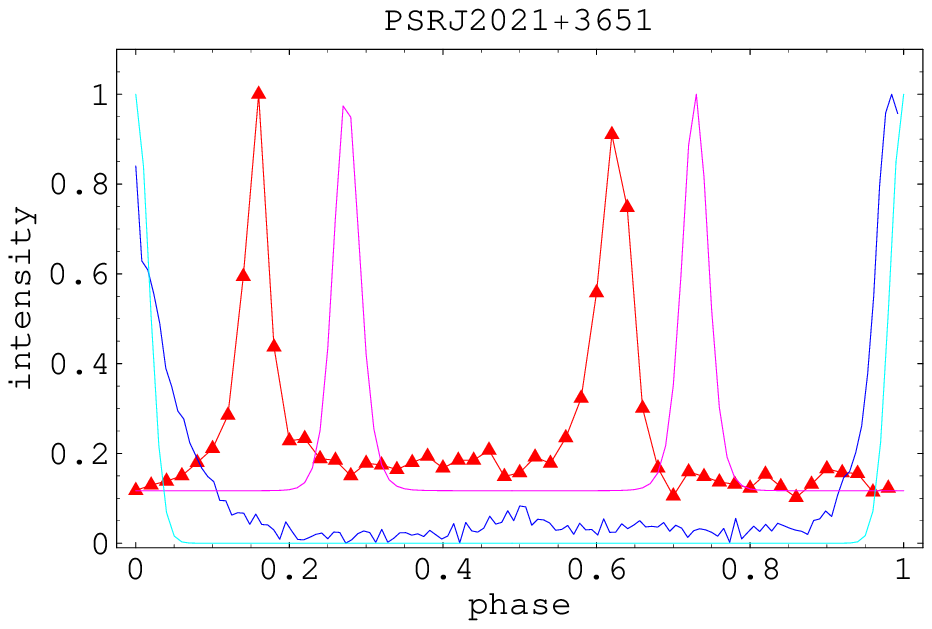} &
  \includegraphics[width=0.45\textwidth]{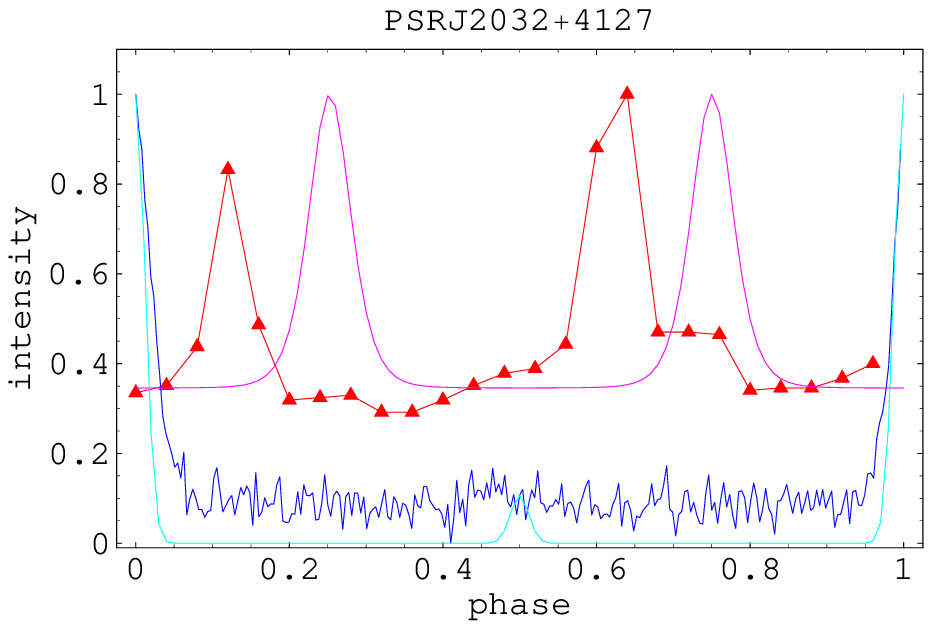} \\
  \includegraphics[width=0.45\textwidth]{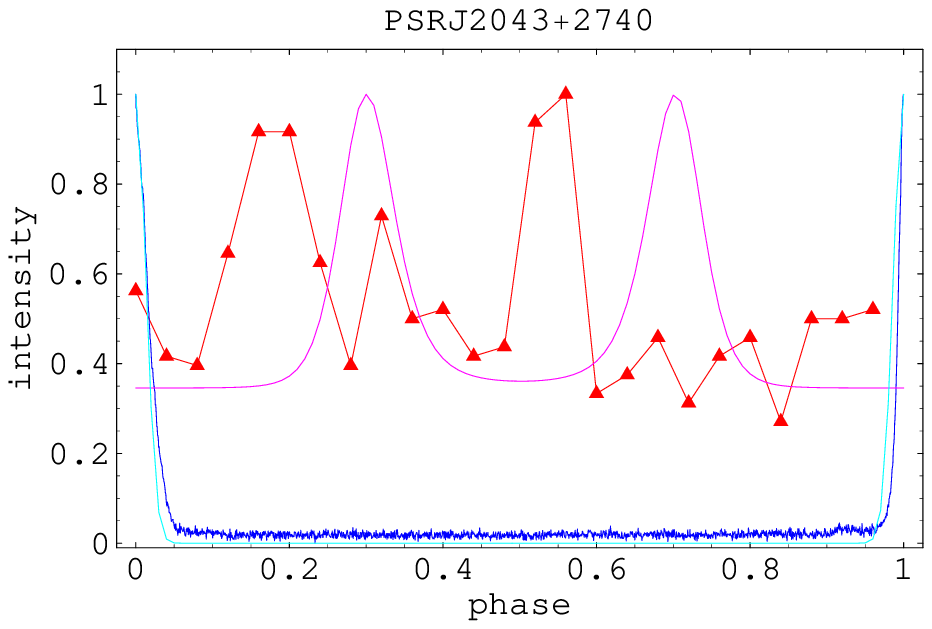} &
  \includegraphics[width=0.45\textwidth]{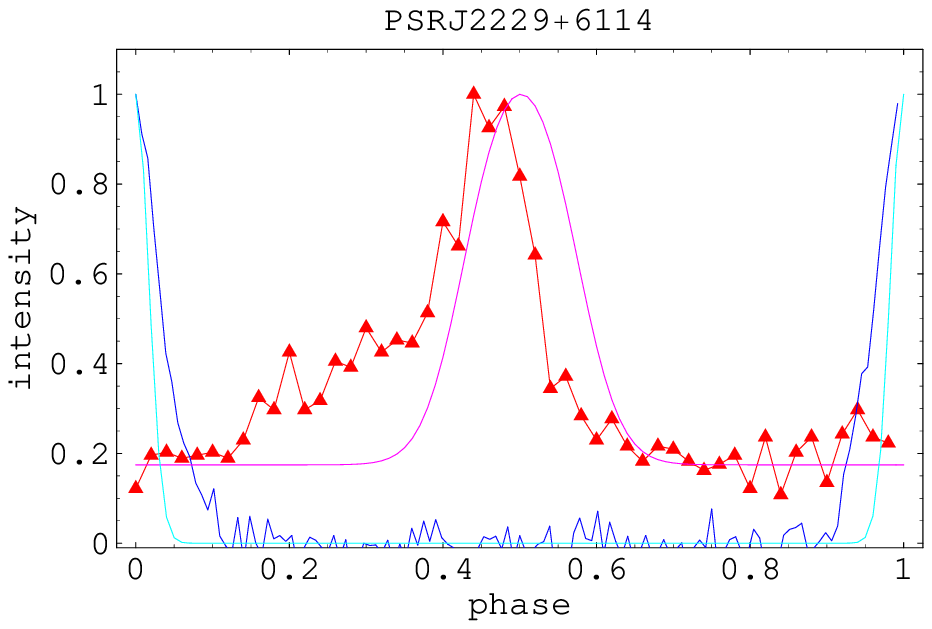} \\
  \end{tabular}
  \caption{Gamma-ray (red lines) and radio (blue lines) light-curves
    of several pulsars fitted with the combined polar cap/striped wind
    model. Intensities are normalized to unity.}
  \label{fig:FermiCL}
\end{figure*}

Finally, PSR J2229+6114 is another example of single pulse gamma-ray
pulsar. The best fit parameters correspond to the same values as for
PSR~J0437-4715. Now, the time lag is even better, again the model
predicts $0.5$ compared to the observed $0.49$,
Fig.~\ref{fig:FermiCL}, bottom right plot.

Details of the parameters used to fit the data are summarized in
Tab.~\ref{tab:Geometrie}.

This ends the sample of fitted gamma-ray pulsars. We demonstrated that
our striped wind/polar cap model implies a well-defined relation
between gamma-rays and radio pulses. Single and double pulse
light-curves are expected. For some pulsars, the model does well in
explaining this relationship. However, it seems that some others pulsars
possess time lag slightly less than the expected value, on average it
is smaller by an amount of $0.1$ in phase. This has probably to find
its root in the geometry of the polar cap or in retardation effects not
taking into account in this investigation.

\subsubsection{Corrected gamma-ray luminosity}

Finally, knowing the precise geometry from the light-curves and time
lag, we computed the flux correction factor for each of the above
mentioned pulsar and summarize our results in
Tab.\ref{tab:Geometrie}. The ``true'' efficiency is also given
therein.

\section{CONCLUSION}
\label{sec:Conclusion}

By combining the striped wind model with a simple prescription for the
polar cap geometry and emission, we were able to derive the time lag
between radio and gamma rays according to the high-energy pulse
separation, if available, in agreement with recent Fermi observations
from the first gamma-ray pulsar catalog. 
According to our composite model, it seems that the observed gamma-ray
pulsation is generated just outside the light-cylinder and radio
pulses come from low altitude polar cap locations.

A further careful inspection of individual pulsar light-curves, the
correlation between their radio and gamma-ray peaks, will permit to
severely constrain the geometry, deducing the obliquity and
inclination of line of sight of the system.

The overestimate of time lag by roughly~0.1 in phase for many pulsars
suggest that another ingredient is still missing to fit properly the
data.  As shown, curved space-time is negligible, so we expect
magnetic field line bending and/or plasma flow within the
magnetosphere and other plasma effects to give us some clue to this
enigma.

\bsp

\label{lastpage}

\end{document}